\documentclass[11pt]{article}
\usepackage{acl}
\usepackage{latexsym}
\usepackage{amsmath}
\usepackage{amssymb}
\usepackage{booktabs}
\usepackage{graphicx}
\usepackage{url}

\usepackage{listings}
\usepackage{enumitem}
\usepackage{float}
\usepackage{needspace}
\usepackage{dblfloatfix}

\usepackage[table]{xcolor}
\usepackage{colortbl}
\usepackage{tcolorbox}

\usepackage{tikz}
\usetikzlibrary{shapes.geometric, arrows, positioning, fit, backgrounds, arrows.meta, calc, shadows}
\usepackage{pgfplots}
\usepgfplotslibrary{groupplots,fillbetween}
\pgfplotsset{compat=1.18}

\definecolor{AethelNavy}{HTML}{0D253D}
\definecolor{AethelIndigo}{HTML}{533AFD}
\definecolor{AethelCream}{HTML}{FDFBF7}

\definecolor{aethelpurple}{RGB}{89,65,242}
\definecolor{hippogreen}{RGB}{112,180,105}
\definecolor{densegray}{RGB}{225,225,225}

\colorlet{aethelNavy}{AethelNavy}
\colorlet{aethelPurple}{AethelIndigo}
\definecolor{aethelBlue}{RGB}{37,99,235}
\definecolor{aethelTeal}{RGB}{20,184,166}
\definecolor{aethelGold}{RGB}{245,158,11}
\colorlet{aethelGreen}{aethelTeal}

\arrayrulecolor{AethelNavy}

\title{Aethel: A Reproducible Graph-Retrieval Framework for Multi-Hop Financial Diligence}

\author{Krish Sapru \\
  \texttt{krish.sapru.th@dartmouth.edu} \\}

\date{}

\begin{document}

\maketitle

\begin{abstract}

Secondary private equity (PE) transactions require rapid synthesis of fragmented, unstructured financial disclosures, where critical metrics and their entity anchors are distributed across disjoint documents with negligible lexical overlap. We present \textbf{Aethel}, a framework that pairs a bipartite Personalized PageRank (PPR) graph-retrieval engine inspired by HippoRAG v2 \cite{gutierrez2025hipporag2} with an orchestrated agent swarm, modelling corpora as bipartite entity-passage graphs and using PPR walks \cite{page1999pagerank} to resolve relational multi-hop queries.

We evaluate Aethel in two complementary settings. First, on 200-question random samples from the official MuSiQue and 2WikiMultiHopQA validation splits under a \emph{closed-set} candidate pool ($\sim$10--20 paragraphs per question), our coreference-aware Bipartite Coreference Teleportation (BCT) layer improves hit-rate coverage over vanilla graph retrieval (HR@5 88.5\% vs.\ 87.5\% on MuSiQue; 100.0\% vs.\ 99.5\% on 2WikiMultiHopQA) at the cost of precision@1---a tradeoff appropriate for multi-passage diligence, where downstream agents require all supporting evidence, not merely the top-ranked passage. Second, and in contrast, we report an \emph{open-corpus} evaluation on a 4{,}123-chunk corpus of real financial disclosures (SEC filings, earnings materials, and secondary-market reports) with 40 manually annotated queries. Here the picture is different: a strong lexical baseline (BM25) outperforms all learned and graph-based methods, while the graph retriever, once supplied with a clean entity index, recovers multi-hop associative recall that a dense bi-encoder misses (multi-hop HR@5 0.600 vs.\ 0.450) but does not overtake BM25 (0.700). Reciprocal Rank Fusion of BM25 and Aethel-NER3 ($k{=}60$, untuned) yields MRR 0.479 vs.\ BM25's 0.468; a paired bootstrap shows this $+$0.011 difference does not clear zero ($N{=}20$, 95\% CI $[-0.163,+0.192]$) and is reported as suggestive rather than significant. We take the gap between the closed-pool and open-corpus results as the paper's central empirical finding: graph-RAG's multi-hop advantage, clearly visible in small candidate pools, is substantially attenuated at open-corpus scale on keyword-dense financial text. Closed-pool results are computed by \texttt{public\_benchmark.py} and open-corpus results by \texttt{evaluate\_aethel.py}, both against real data; code is released for reproduction.

\end{abstract}

\section{Introduction}

The underwriting of secondary PE transactions demands factual precision under compressed timelines: analysts cross-reference private LP letters, financial workbooks, capitalization structures, and public SEC filings to evaluate asset values, unfunded liabilities, and fund rules. This requires resolving cross-document, multi-hop dependencies. Standard Dense RAG \cite{lewis2020rag, karpukhin2020dpr} maps chunks into an embedding space and ranks by cosine similarity; effective for single-hop queries, dense bi-encoders falter when a metric and its anchoring entity are topographically isolated. For example, when a portfolio company is named on one page of an asset ledger and the fund's aggregate liabilities appear a dozen pages later, the two segments share little lexical or contextual overlap, so a dense index displaces the bridging context and the downstream LLM hallucinates.

To handle these relational gaps, recent literature has shifted toward Graph-Augmented RAG (GraphRAG) systems \cite{edge2024graphrag, peng2024graphrag}. Neurobiologically inspired frameworks like HippoRAG \cite{gutierrez2024hipporag, gutierrez2025hipporag2} deploy Personalized PageRank (PPR) \cite{page1999pagerank} over bipartite entity-passage configurations to resolve complex multi-hop queries without relying on expensive global graph clustering.
\begin{figure}[t]
\centering
\begin{tikzpicture}[
    font=\scriptsize,
    box/.style={
        rectangle,
        rounded corners=3pt,
        draw=black!35,
        thick,
        align=center,
        minimum width=2.25cm,
        minimum height=0.85cm,
        inner sep=2pt
    },
    arrow/.style={
        -{Latex[length=1.6mm]},
        thick,
        draw=black!60
    }
]

\node[box, fill=blue!8] (docs) at (-3.35,0) {
    \textbf{Docs}\\[-1pt]
    \tiny PDFs, CSVs, Excel
};

\node[box, fill=blue!8] (ingest) at (-0.655,0) {
    \textbf{Ingest}\\[-1pt]
    \tiny text extraction  
};

\node[box, fill=aethelPurple!18] (graph) at (2.10,0) {
    \textbf{Graph}\\[-1pt]
    \tiny entities + passages
};

\node[box, fill=aethelPurple!25] (orch) at (-3.35,-1.35) {
    \textbf{Orchestrator}\\[-1pt]
    \tiny routes evidence
};

\node[box, fill=aethelGreen!22] (agents) at (-0.655,-1.35) {
    \textbf{Agents}\\[-1pt]
    \tiny valuation, leverage
};

\node[box, fill=aethelGold!35] (memo) at (2.0,-1.35) {
    \textbf{IC Memo}\\[-1pt]
    \tiny risks + metrics
};

\draw[arrow] (docs.east) -- (ingest.west);
\draw[arrow] (ingest.east) -- (graph.west);

\draw[arrow] (graph.south) -- ++(0,-0.28) -| (orch.north);
\draw[arrow] (orch.east) -- (agents.west);
\draw[arrow] (agents.east) -- (memo.west);

\end{tikzpicture}

\vspace{-0.5em}
\caption{Aethel converts financial documents into graph-indexed evidence, routes it through specialist diligence agents, and synthesizes an investment committee memo.}
\label{fig:aethel_pipeline}
\vspace{-0.8em}
\end{figure}
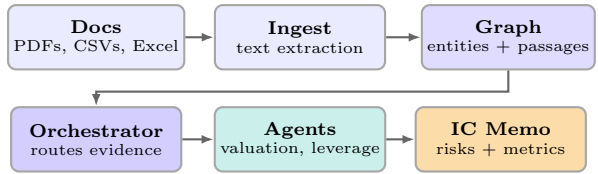
However, existing multi-agent and graph frameworks remain brittle when deployed in document-intensive corporate diligence domains, where critical financial metrics are topographically isolated across disjoint documents and entity names appear in multiple surface forms.

We address these limitations by introducing \textbf{Aethel}, a graph-retrieval framework for multi-hop financial diligence. Aethel routes multi-hop queries via sparse power iteration (SpMV) over a bipartite entity-passage graph and forwards the retrieved evidence to a downstream specialist swarm. It prioritizes hit rate---the probability that all necessary supporting passages appear in the top-$K$---over top-1 precision, since a specialist can filter irrelevant context but is blocked when a bridge passage is missing, and its graph topology yields interpretable, auditable entity-passage traversal paths.

To summarize our core contributions:
\begin{itemize}[noitemsep,topsep=0pt,parsep=0pt,partopsep=0pt]
    \item \textbf{Bipartite Coreference Teleportation (BCT)}: We formalize a coreference correction layer that expands entity mentions in the PPR personalization vector via alias matching and substring overlap, improving HR@5 (and HR@3 on 2WikiMultiHopQA) at the cost of precision@1---a tradeoff that benefits multi-passage diligence workflows where all supporting evidence must be surfaced.
    \item \textbf{Graph-Filtered Specialist Swarm (system description)}: We describe a system architecture coupling bipartite PPR retrieval with a parallel agent swarm, in which the graph-walk step acts as a topological filter, constraining the context window forwarded to each specialist to only the highest-relevance passages and preserving cross-document audit trails. We present this as an architecture rather than a validated result: agent-level performance is not separately benchmarked in this work, and specialist swarm evaluation is left as future work.
    \item \textbf{An open-corpus financial evaluation and a scale-dependence finding}: We construct a 4{,}123-chunk corpus of real financial disclosures with 40 manually annotated queries and show that graph-RAG's multi-hop advantage over dense retrieval, visible in closed candidate pools, does \emph{not} extend to beating a strong lexical baseline (BM25) at open-corpus scale on keyword-dense financial text. An explicit Reciprocal Rank Fusion of the two leaves BM25's recall lead intact and yields only a small MRR difference that is not statistically distinguishable from zero at this sample size. We report this gap honestly and diagnose its cause. All evaluation code is released; results are deterministically reproducible.
\end{itemize}

\section{Related Work}
\label{sec:related_work}

\subsection{Graph-Augmented Retrieval}
Dense RAG frameworks rely on isolated semantic embeddings, which break down on queries spanning disjoint documents or requiring topological reasoning \cite{lewis2020rag}; a common remedy injects structured knowledge graphs into the retrieval loop \cite{karpukhin2020dpr}. GraphRAG synthesizes global community summaries across hierarchical clusters \cite{edge2024graphrag} but incurs heavy token overhead, whereas HippoRAG runs personalized PageRank over sparse bipartite entity-passage graphs, avoiding costly global summaries \cite{gutierrez2025hipporag2}. HybridRAG \cite{sarmah2024hybridrag} combines knowledge graphs and vector retrieval for financial and enterprise QA. Our open-corpus finding stands in partial tension with HybridRAG's reported gains: where they report a graph--vector combination helping on financial QA, we find the graph component does not beat BM25 alone at open-corpus scale, which we attribute to corpus size and query keyword density. \textit{Aethel} adapts this graph-walk paradigm to noisy, high-density financial documents with complex tabular ledgers.

\subsection{Multi-Agent Collaboration and Swarms}
Parallel work substitutes monolithic models with specialized multi-agent orchestrations \cite{autogen} that divide a problem across independently-prompted components. These topologies perform well on software-engineering and math-reasoning tasks, but extending them to long-form financial diligence remains open: agents operating over standard vector indexes stall when cross-document references are split or corrupted during extraction. Aethel instead maps its specialists over a bipartite entity-graph layer, so they collaborate under hard structural constraints that resist contextual fragmentation.

\section{System Architecture \& The Specialist Swarm}

Unlike flat, linear pipelines, Aethel decouples macro-document retrieval, deep external validation, and narrative synthesis into an orchestrated multi-agent layout \cite{autogen}. We describe this architecture for completeness; its agent-level behavior is not separately benchmarked, and the quantitative claims in this paper concern only the retrieval layer. A central Orchestrator decomposes multi-faceted financial queries into atomic sub-tasks, so that each specialist receives a narrow, well-defined context rather than a monolithic prompt.

The swarm comprises specialist nodes over the graph-retrieved evidence: a \textbf{Liquidity Specialist} (capital-call liabilities, unfunded obligations, distribution pacing), a \textbf{Valuation Auditor} (NAV methodologies and Level III inputs), a \textbf{Diligence Auditor} (compliance friction and line-item contradictions between public and private reports), and a \textbf{Market Research Specialist} (live public comps via an external web-search API, not part of the quantitative evaluation). A \textbf{Synthesizer} resolves discrepancies via an internal consensus protocol and compiles the final memo.

\section{Mathematical Multi-Hop Framework}

To formalize how Aethel bypasses the lexical boundaries of conventional vector search, we define our graph retrieval architecture using standard bipartite graph notation.

\subsection{Bipartite Entity-Passage Graph Representation}

Let an ingested document corpus be represented as an undirected bipartite graph:
\begin{equation}
\mathcal{G} = (\mathcal{V}_P \cup \mathcal{V}_E, \mathcal{E})
\end{equation}
Where $\mathcal{V}_P = \{p_1, p_2, \dots, p_N\}$ represents discrete text passages (document chunks), $\mathcal{V}_E = \{e_1, e_2, \dots, e_M\}$ represents unique extracted entity nodes, and $\mathcal{E}$ is the set of edges mapping undirected co-occurrence. An edge $(p_i, e_j) \in \mathcal{E}$ exists if and only if entity $e_j$ was extracted from passage $p_i$ during indexing.

\subsection{Transition Probability Matrix}

Let $\mathbf{A} \in \{0, 1\}^{(N+M) \times (N+M)}$ be the adjacency matrix of $\mathcal{G}$. The transition probability matrix $\mathbf{W} \in \mathbb{R}^{(N+M) \times (N+M)}$ dictates a random walk sequence across the bipartite topology, defined by column-normalizing $\mathbf{A}$:
\begin{equation}
W_{ij} = \frac{A_{ij}}{\sum_{k} A_{kj}}
\end{equation}
For sink nodes with no outgoing edges, their columns in $\mathbf{W}$ are set to zero.

\subsection{Personalized PageRank (PPR) Retrieval}

During user query execution, a query $q$ is parsed to extract query entities
$\mathcal{E}_q \subseteq \mathcal{V}_E$. These entities serve as seed nodes in
the bipartite entity--passage graph. Rather than relying only on direct dense
similarity, PPR allows relevance to propagate through shared entities, linked
passages, and intermediate relational bridges.

We construct a personalization vector $\mathbf{s} \in \mathbb{R}^{N+M}$
concentrated on the active seed nodes:
\begin{equation}
s_i =
\begin{cases}
\frac{1}{|\mathcal{E}_q|}, & \text{if node } i \in \mathcal{E}_q \\
0, & \text{otherwise}.
\end{cases}
\end{equation}

Let $\mathbf{W} \in \mathbb{R}^{(N+M)\times(N+M)}$ denote the normalized graph
transition matrix, where $N$ is the number of entity nodes and $M$ is the number
of passage nodes. The steady-state probability vector
$\mathbf{p} \in \mathbb{R}^{N+M}$ is solved using power iteration:
\begin{equation}
\mathbf{p}^{(t+1)} = \alpha \mathbf{s} + (1 - \alpha)\mathbf{W}\mathbf{p}^{(t)}.
\end{equation}

Here, $\alpha = 0.15$ is the damping factor. Intuitively, the teleportation term
keeps the walk anchored to the query entities, while the transition term allows
evidence to diffuse across neighboring passages and related entities. This is
important for financial diligence, where relevant evidence is often distributed
across separate documents, tables, covenants, disclosures, and management notes.

Power iteration continues until the ranking stabilizes or a fixed iteration
cap is reached.  Theoretically, convergence is guaranteed when:
\begin{equation}
\left\|\mathbf{p}^{(t+1)} - \mathbf{p}^{(t)}\right\|_1 < \epsilon .
\end{equation}
In our implementation, we cap iterations at $I = 20$, which is sufficient for
convergence given $\alpha = 0.15$ and the sparse graph structure.

Upon convergence, all passage nodes $p_i \in \mathcal{V}_P$ are ranked according
to their steady-state probability masses in $\mathbf{p}$. The retrieved context
is therefore given by:
\begin{equation}
\mathrm{TopK}(q) =
\operatorname*{arg\,topK}_{p_i \in \mathcal{V}_P} \mathbf{p}_{p_i}.
\end{equation}

The top-$K$ passages are then passed to the downstream agentic swarm along with
their source metadata, linked entities, and PPR scores. This provides both
retrieval evidence and a lightweight explanation of why each passage was selected.
In practice, this improves retrieval for multi-hop queries because passages can
be recovered through graph proximity even when paraphrasing or surface-form
variation weakens direct lexical matching.

\Needspace{18\baselineskip}
\begin{figure}[H]
\centering
\begin{tikzpicture}[
  font=\footnotesize,
  node distance=8mm and 14mm,
  >=Stealth,
  ent/.style={circle, draw=green!70!black, fill=green!12, minimum size=11mm, inner sep=0pt},
  pas/.style={rectangle, draw=orange!80!black, fill=orange!12, rounded corners=3pt, minimum width=3.55cm, minimum height=10mm, text width=3.45cm, align=center},
  query/.style={rectangle, draw=blue!70!black, fill=blue!10, rounded corners=3pt, minimum width=4.25cm, minimum height=10mm, align=center, font=\bfseries\footnotesize},
  op/.style={rectangle, draw=purple!70!black, fill=purple!10, rounded corners=3pt, minimum width=4.25cm, minimum height=10mm, align=center, font=\bfseries\footnotesize},
  edge/.style={draw=gray!55, line width=0.8pt},
  hi/.style={draw=red!75!black, line width=1.4pt},
  arrow/.style={->, draw=gray!80, line width=0.95pt},
  pill/.style={fill=white, draw=gray!30, rounded corners=2pt, inner sep=1.2pt, font=\scriptsize, text=gray!70}
]

\node (q)   [query] {Multi-hop Query};
\node (ppr) [op, below=of q] {PPR / SpMV Iteration};

\node (e1) [ent, hi, below left=6mm and 14mm of ppr] {$e_1$};
\node (e2) [ent, below=of e1] {$e_2$};
\node (e3) [ent, hi, below=of e2] {$e_3$};

\node (p1) [pas, hi, right=26mm of e1] {Passage $p_1$\\cap table};
\node (p2) [pas, below=of p1] {Passage $p_2$\\NAV note};
\node (p3) [pas, hi, below=of p2] {Passage $p_3$\\audit memo};

\draw[edge, hi] (e1) to[bend left=12] (p1);
\draw[edge] (e1) -- (p2);
\draw[edge] (e2) -- (p2);
\draw[edge] (e2) to[bend right=8] (p3);
\draw[edge, hi] (e3) to[bend right=-8] (p1);
\draw[edge, hi] (e3) -- (p3);

\draw[arrow] (q) -- node[pill, pos=0.40, right=-8pt] {seed} (ppr);

\draw[arrow] (ppr.west) -- ++(-6mm,0) |- node[pill, pos=0.25] {$\mathbf{s}$} (e1.east);

\node (topk) [query, below=12mm of p3] {Top-$K$ Passages};
\draw (p3.south) -- ++(0,-4mm) |- node[pill, pos=0.06] {rank by $\mathbf{p}$} (topk.north);

\node[font=\scriptsize, text=gray!70, below=1mm of topk, text width=5.5cm, align=center]
  {red = example 2-hop retrieval path:\\$e_1 \!\to\! p_1 \!\to\! e_3 \!\to\! p_3$};

\end{tikzpicture}
\caption{Bipartite entity--passage retrieval in Aethel.  Query entities seed the PPR personalization vector~$\mathbf{s}$; power iteration diffuses probability mass through entity--passage edges (no passage--passage edges exist).  The red path illustrates a two-hop chain: seed entity~$e_1$ activates passage~$p_1$, which shares entity~$e_3$, propagating mass onward to passage~$p_3$.  All passages are ranked by their steady-state probability~$\mathbf{p}$ and the top-$K$ are forwarded to downstream agents.}
\label{fig:bipartite_ppr}
\end{figure}
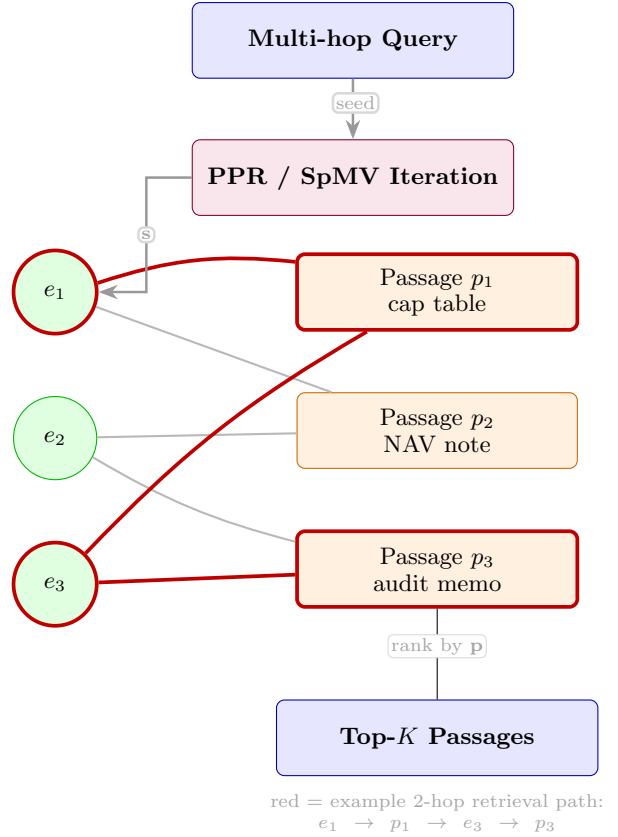

\subsection{Scalability Analysis \& PPR Complexity}

Entity extraction is $\mathcal{O}(N)$ and the SpMV power iteration scales as $\mathcal{O}(I \cdot |\mathcal{E}|)$; with $\alpha = 0.15$ and $I = 20$ over a sparse graph, iteration converges quickly and avoids the super-linear cost of global graph re-clustering. We note, however, that favorable convergence complexity does not translate into retrieval quality at open-corpus scale; Section~\ref{sec:open_corpus} shows the graph retriever's precision degrades as the passage pool grows.

\section{Experimental Setup \& Compliance}

\subsection{Closed-Pool Evaluation Protocol}

We evaluate Aethel on 200-question random samples (seed=42) from the official MuSiQue and 2WikiMultiHopQA validation splits. For each question, the provided context paragraphs (approximately 10 paragraphs for 2WikiMultiHopQA and 20 paragraphs for MuSiQue, representing gold supporting paragraphs mixed with distractors) are indexed into fresh per-question retrieval indices. We compare sparse lexical (TF-IDF), dense bi-encoder (all-MiniLM-L6-v2), exact bipartite PPR graph, and coreference-aware BCT retrieval. Importantly, retrieval is conducted only over this closed context pool for each question, rather than a full open corpus. We evaluate passage selection via Hit Rate@$k$ (HR@$k$) and Mean Reciprocal Rank (MRR) against gold answers. No query template, hyperparameter, or system component was tuned on these datasets.

\subsection{Open-Corpus Evaluation Protocol}

To test retrieval under production-realistic conditions, we additionally construct an open-corpus benchmark of 4{,}123 chunks drawn from real financial disclosures: SEC 10-K filings (NVIDIA, Salesforce), asset-manager earnings materials (Blackstone, Apollo), and secondary-market advisory reports (Jefferies), together with three synthetic fund documents authored to mirror LP quarterly-reporting conventions. We manually annotate 40 queries (20 single-hop, 20 multi-hop), freezing gold passage labels \emph{before} any retriever is executed. All systems retrieve over the identical 4{,}123-chunk index. We compare BM25, a dense bi-encoder (all-MiniLM-L6-v2), the regex-graph retriever (Aethel-Reg), a filtered-NER graph retriever with substring adjacency (Aethel-NER3), and a Reciprocal Rank Fusion of BM25 and Aethel-NER3 (Hybrid-RRF). This evaluation is preliminary: gold labels are single-annotator, and we report effect sizes together with bootstrap confidence intervals rather than relying on significance thresholds.

\subsection{Reproducibility and Benchmark Protocol}

All closed-pool benchmark evaluations use the \texttt{datasets} library to stream the official HuggingFace validation splits of MuSiQue and 2WikiMultiHopQA. For benchmark reproducibility and to avoid external API dependencies, \texttt{public\_benchmark.py} implements the PPR retrieval and BCT layers as a self-contained evaluator using numpy sparse matrix-vector multiplication (SpMV). This standalone evaluator implements the same bipartite PPR algorithm described in Section~4 with regex-based entity extraction, operating on each question's provided context paragraphs. The open-corpus financial evaluation is implemented in \texttt{evaluate\_aethel.py}. The production diligence pipeline (\texttt{rag\_service.py}) delegates to the HippoRAG v2 library for LLM-based entity extraction and graph construction; BCT integration into the production pipeline is left as future work.

For each closed-pool question we build a fresh per-question bipartite entity--passage graph, a sparse TF-IDF index, and a dense bi-encoder index (using a locally run \texttt{all-MiniLM-L6-v2} model) over the provided context paragraphs. Results are cached locally to \texttt{eval\_cache.json} for deterministic reproduction on re-runs. Because PPR here is a deterministic power iteration over a fixed graph and seed vector, re-runs are bit-identical. The complete evaluation code is released at \texttt{backend/public\_benchmark.py} and \texttt{backend/evaluate\_aethel.py}. The benchmark script additionally emits a Retrieval Token Overlap (RTO) score---token-level F1 between the concatenation of retrieved passages and the gold answer string. We omit RTO from Table~\ref{tab:musique_2wiki} because, computed over full passages, it conflates retrieval relevance with passage length and is not comparable across systems with different top-$K$ budgets; HR@$k$ and MRR characterize retrieval quality without this confound.

\section{Empirical Results \& Discussion}

We report (i) closed-pool benchmark performance on established multi-hop QA datasets, (ii) an open-corpus evaluation on real financial documents, and (iii) a precision--coverage tradeoff analysis.

\subsection{Domain Bridge: Mapping General QA to Financial Diligence}

We evaluate on two established multi-hop QA benchmarks as a proxy for cross-document retrieval capability, and complement them with the direct open-corpus financial evaluation in Section~\ref{sec:open_corpus}, which is the more faithful test. MuSiQue \cite{trivedi2022musique} stresses compositional retrieval---each question requires a connected chain of evidence, so a single missing bridge passage breaks the answer---mirroring PE diligence, where a missing link in a Portfolio Company $\rightarrow$ Ledger $\rightarrow$ Fund $\rightarrow$ Unfunded-Liability chain invalidates the analysis. 2WikiMultiHopQA \cite{ho2020constructing} stresses multi-document coreference across dispersed files, analogous to tracking capital accounts across LP letters, GP disclosures, and public filings. We treat multi-hop capability on these sets as necessary but not sufficient for diligence retrieval.

\subsection{Closed-Pool Benchmark Evaluation (MuSiQue, 2WikiMultiHopQA)}
\label{sec:musique_2wiki}

Table~\ref{tab:musique_2wiki} reports the retrieval metric suite on 200-question random samples from both benchmarks. We report HR@1, HR@3, HR@5, and MRR. Importantly, because the evaluation is performed over each question's closed context pool ($\sim$10--20 paragraphs), the dense bi-encoder baseline (MiniLM) achieves very high absolute performance (e.g., 96.0\% HR@5 on MuSiQue). This is expected since dense embeddings are highly effective at short-list ranking where the distraction pool is small and open-domain false positives are absent.

Comparing the graph-based retrievers, we observe a clear \textbf{precision--coverage tradeoff} introduced by Aethel's BCT layer: BCT improves graph retrieval coverage (HR@5 from 87.5\% to 88.5\% on MuSiQue, and from 99.5\% to 100.0\% on 2WikiMultiHopQA) at the cost of precision@1 (HR@1: 57.0\% vs.~63.0\% on MuSiQue). This is a favorable tradeoff for multi-passage diligence workflows, where downstream specialist agents require \emph{all} gold supporting passages in the context window to perform multi-hop synthesis and numerical auditing---not merely the single top-ranked passage. We stress that these are \emph{closed-pool} results; Section~\ref{sec:open_corpus} shows they do not carry over to open-corpus retrieval.

\begin{table*}[t]
\centering
\small
\setlength{\tabcolsep}{5.5pt}
\renewcommand{\arraystretch}{1.15}
\begin{tabular}{@{}lccccccccc@{}}
\toprule
\textbf{Method} & \multicolumn{4}{c}{\textbf{2WikiMultiHopQA}} & & \multicolumn{4}{c}{\textbf{MuSiQue}} \\
\cmidrule(lr){2-5} \cmidrule(lr){7-10}
 & \textbf{HR@1} & \textbf{HR@3} & \textbf{HR@5} & \textbf{MRR} & & \textbf{HR@1} & \textbf{HR@3} & \textbf{HR@5} & \textbf{MRR} \\
\midrule
Sparse (TF-IDF) & 79.0 & 96.5 & 99.0 & 0.873 & & 58.0 & 78.0 & 84.5 & 0.681 \\
Dense (MiniLM)  & \textbf{90.0} & \textbf{99.0} & 99.5 & \textbf{0.940} & & \textbf{76.5} & \textbf{93.5} & \textbf{96.0} & \textbf{0.849} \\
Graph (PPR)     & 83.0 & 97.0 & 99.5 & 0.900 & & 63.0 & 80.0 & 87.5 & 0.721 \\
\textbf{Aethel (PPR+BCT)} & 78.5 & 98.0 & \textbf{100.0} & 0.877 & & 57.0 & 78.5 & 88.5 & 0.687 \\
\bottomrule
\end{tabular}
\caption{Closed-pool retrieval metrics on 200-question random samples (seed=42) from the official MuSiQue and 2WikiMultiHopQA validation splits. Retrieval is conducted over each question's provided context pool ($\sim$10--20 paragraphs), not open-corpus. HR@$k$ = Hit Rate (\%). MRR = Mean Reciprocal Rank. \textbf{Bold} = best per column. BCT improves hit rate coverage of the graph-based retriever at the cost of precision@1. All numbers computed by \texttt{public\_benchmark.py} against real HuggingFace data.}
\label{tab:musique_2wiki}
\end{table*}

\subsection{Open-Corpus Financial Retrieval}
\label{sec:open_corpus}

The benchmarks in Section~\ref{sec:musique_2wiki} evaluate retrieval over each question's provided pool of $\sim$10--20 paragraphs. Production diligence, however, is an \emph{open-corpus} problem: the relevant passages must be located among thousands of chunks with no per-question candidate set. Table~\ref{tab:financial_eval} reports retrieval over our 4{,}123-chunk financial corpus. Four findings stand out.

First, \textbf{BM25 is the strongest method overall on HR@5}, reflecting the proper-noun density of financial queries (``Fee-Related Earnings,'' ``Strategic Partners,'' ``NVIDIA Corporation''), where exact lexical matching is difficult to beat. Second, \textbf{the graph retriever's advantage over dense retrieval does not extend to BM25}: Aethel-NER3 overtakes the dense bi-encoder on multi-hop (HR@5 0.600 vs.\ 0.450) but not BM25 (HR@5 0.700). Third, \textbf{Reciprocal Rank Fusion (BM25 $\oplus$ Aethel-NER3) matches BM25's HR@1 on multi-hop and produces a higher MRR (0.479 vs.\ 0.468)}, but this $+$0.011 difference does not clear zero under a paired bootstrap (95\% CI $[-0.163,+0.192]$, $N{=}20$; see Table~\ref{tab:financial_eval}). The fusion does not lift HR@5 above BM25 (0.650 vs.\ 0.700), so the overall picture remains that BM25 dominates at recall, and any graph contribution at the top of the ranking is small and, at this sample size, not statistically distinguishable from zero. Fourth, \textbf{entity-index quality is decisive}: the original regex graph (Aethel-Reg) trails substantially (multi-hop HR@5 0.400), and an intermediate naive-spaCy variant failed almost entirely due to a context-dependent adjacency defect---edges were created only where the NER model re-extracted an entity's exact surface form, so ``NorthRiver'' and ``NorthRiver IV LP $\ldots$'' became disconnected nodes and propagation collapsed. Resolving this with substring adjacency over a filtered entity set (dropping $\sim$20k OCR fragments and bare numerals from $\sim$27k raw extractions, leaving $\sim$6k clean entities) recovered the reported results.

We read the contrast with Section~\ref{sec:musique_2wiki} directly, and take it as the paper's central empirical finding: the graph retriever's clean multi-hop wins under a 10--20 paragraph closed pool do not survive the transition to a 4{,}123-chunk open corpus, where BM25 leads on HR@5 and fusion yields at most a small MRR difference that we cannot distinguish from zero. PPR's teleport-and-diffuse dynamics, which isolate the bridge passage cleanly in a small pool, spread mass across a large heterogeneous graph, while a well-tuned lexical baseline recovers the keyword-anchored answers more reliably. The RRF result indicates the graph's residual contribution is, at most, narrow; at $N{=}20$ it is not statistically distinguishable from no effect.

\paragraph{Qualitative example.} For the query \emph{``How do Blackstone and Apollo each characterize secondaries activity in their most recent results?''}, BM25 and the dense baseline anchor on Blackstone chunks and miss the Apollo side, whereas Aethel-NER3 reaches the Apollo earnings passage by propagating from the shared \textsc{secondaries} entity---the kind of cross-document recall a graph provides but flat retrievers miss. The aggregate table shows this mechanism does not translate into a statistically distinguishable gain on this corpus.

\begin{table}[t]
\centering
\small
\setlength{\tabcolsep}{2pt}
\renewcommand{\arraystretch}{0.95}
\begin{tabular}{@{}llcccc@{}}
\toprule
\textbf{System} & \textbf{Group} & \textbf{HR@1} & \textbf{HR@5} & \textbf{MRR} & \textbf{R@5} \\
\midrule
BM25            & Multi & 0.350 & \textbf{0.700} & 0.468 & \textbf{0.525} \\
Dense           & Multi & 0.200 & 0.450 & 0.318 & 0.275 \\
Aethel-Reg      & Multi & 0.050 & 0.400 & 0.175 & 0.200 \\
Aethel-NER3     & Multi & 0.150 & 0.600 & 0.351 & 0.350 \\
\textbf{Hybrid-RRF} & Multi & \textbf{0.350} & 0.650 & \textbf{0.479} & 0.475 \\
\midrule
BM25            & All & \textbf{0.225} & \textbf{0.650} & \textbf{0.367} & \textbf{0.562} \\
Dense           & All & 0.200 & 0.450 & 0.311 & 0.362 \\
Aethel-Reg      & All & 0.025 & 0.275 & 0.121 & 0.175 \\
Aethel-NER3     & All & 0.075 & 0.425 & 0.226 & 0.300 \\
\textbf{Hybrid-RRF} & All & \textbf{0.250} & 0.475 & 0.348 & 0.388 \\
\bottomrule
\end{tabular}
\caption{Open-corpus retrieval on 40 annotated queries (20 single-hop, 20 multi-hop) over a 4{,}123-chunk financial corpus. All systems retrieve over the identical index; gold passages frozen pre-retrieval; single annotator; preliminary. Hybrid-RRF fuses BM25 and Aethel-NER3 ranked lists (top-100 each) via Reciprocal Rank Fusion ($k{=}60$, the canonical untuned value from \cite{cormack2009reciprocal}); $k$ was fixed before inspecting output. Hybrid-RRF matches BM25 HR@1 on multi-hop and produces an MRR of 0.479 vs.\ BM25's 0.468 ($+$0.011); a paired bootstrap 95\% CI on the MRR difference is $[-0.163, +0.192]$ ($N{=}20$), so this gain does not clear zero and is reported as suggestive, not significant. \textbf{Bold} = best per row-group.}
\label{tab:financial_eval}
\end{table}

\subsection{Scale Sensitivity}
\label{sec:scaling}

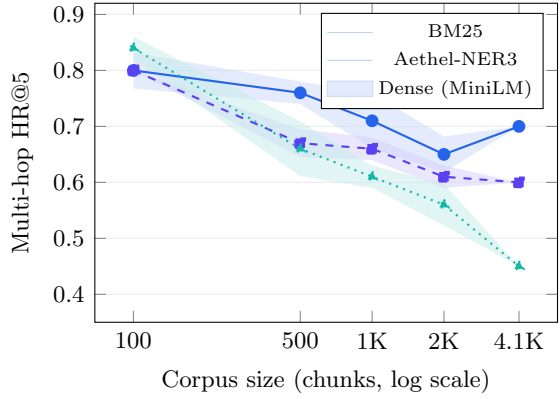
\begin{figure}[t]
\centering
\begin{tikzpicture}
\begin{axis}[
    width=\columnwidth,
    height=5.8cm,
    xmode=log,
    log basis x=10,
    xtick={100,500,1000,2000,4123},
    xticklabels={100,500,1K,2K,4.1K},
    xlabel={Corpus size (chunks, log scale)},
    ylabel={Multi-hop HR@5},
    ymin=0.35, ymax=0.92,
    ytick={0.4,0.5,0.6,0.7,0.8,0.9},
    ymajorgrids=true,
    grid style={gray!15},
    legend pos=north east,
    legend style={font=\scriptsize, inner sep=2pt},
    tick label style={font=\small},
    label style={font=\small},
    clip=false,
]
\addplot[aethelBlue!30, name path=bm25hi, draw=none]
    coordinates {(100,0.832)(500,0.780)(1000,0.759)(2000,0.682)(4123,0.700)};
\addplot[aethelBlue!30, name path=bm25lo, draw=none]
    coordinates {(100,0.768)(500,0.740)(1000,0.661)(2000,0.618)(4123,0.700)};
\addplot[aethelBlue!25, fill opacity=0.5] fill between[of=bm25hi and bm25lo];
\addplot[aethelBlue, thick, solid, mark=*, mark size=2pt]
    coordinates {(100,0.800)(500,0.760)(1000,0.710)(2000,0.650)(4123,0.700)};
\addlegendentry{BM25}
\addplot[aethelpurple!30, name path=ner3hi, draw=none]
    coordinates {(100,0.800)(500,0.695)(1000,0.680)(2000,0.630)(4123,0.600)};
\addplot[aethelpurple!30, name path=ner3lo, draw=none]
    coordinates {(100,0.800)(500,0.645)(1000,0.640)(2000,0.590)(4123,0.600)};
\addplot[aethelpurple!25, fill opacity=0.5] fill between[of=ner3hi and ner3lo];
\addplot[aethelpurple, thick, dashed, mark=square*, mark size=1.8pt]
    coordinates {(100,0.800)(500,0.670)(1000,0.660)(2000,0.610)(4123,0.600)};
\addlegendentry{Aethel-NER3}
\addplot[aethelTeal!30, name path=densehi, draw=none]
    coordinates {(100,0.860)(500,0.709)(1000,0.630)(2000,0.597)(4123,0.450)};
\addplot[aethelTeal!30, name path=denselo, draw=none]
    coordinates {(100,0.820)(500,0.611)(1000,0.590)(2000,0.523)(4123,0.450)};
\addplot[aethelTeal!25, fill opacity=0.5] fill between[of=densehi and denselo];
\addplot[aethelTeal, thick, dotted, mark=triangle*, mark size=2pt]
    coordinates {(100,0.840)(500,0.660)(1000,0.610)(2000,0.560)(4123,0.450)};
\addlegendentry{Dense (MiniLM)}
\end{axis}
\end{tikzpicture}
\caption{Multi-hop HR@5 as corpus size grows from 100 to 4{,}123 chunks, over the 20 multi-hop queries. Gold passages are retained in every subsample (see text). Bands show $\pm 1\sigma$ over 5 resamples (seed 42); the full-corpus point is a single deterministic trial. Dense retrieval leads at the smallest pool (0.840 at 100 chunks) but collapses to 0.450 at full scale, while BM25 (0.800$\rightarrow$0.700) is robust to distractor growth. The graph retriever overtakes Dense from 500 chunks onward and the Aethel-NER3--Dense gap widens to 0.150 at full corpus (0.600 vs.\ 0.450), though Aethel-NER3 never overtakes BM25. The full-corpus Aethel-NER3 point (0.600) matches its multi-hop HR@5 in Table~\ref{tab:financial_eval}.}
\label{fig:scaling}
\end{figure}

To characterise how each retriever degrades as the corpus grows, we subsample the 4{,}123-chunk financial corpus to five sizes (100, 500, 1k, 2k, 4.1k) and measure multi-hop HR@5 across the 20 multi-hop queries. Gold passages are \emph{always} retained in each subsample so that HR@5 degrades only as distractors accumulate, not because the answer is absent; this isolates retrievability from presence. We use five random resamples per non-full size (seed 42) to estimate variance, and a single deterministic trial at full scale. Figure~\ref{fig:scaling} shows the results.

Three findings emerge. First, \textbf{BM25 leads at every corpus size from 500 chunks upward} (0.760 at 500 to 0.700 at full scale), a mild degradation that reflects its robustness to distractor growth; at the smallest pool it is edged by Dense (0.800 vs.\ 0.840). The graph retriever never overtakes BM25 at any tested scale, so no graph/lexical crossover is supported by the data. Second, \textbf{Dense retrieval leads at the smallest pool but collapses at scale}: at 100 chunks Dense achieves HR@5 0.840, the highest of any system, but by 4{,}123 chunks it falls to 0.450---a drop of 0.390. As passages accumulate the embedding space becomes underdiscriminative, and cosine similarity loses the ability to separate the target passage from semantically similar distractors. Third, \textbf{the graph retriever overtakes Dense as the corpus grows}: the two are level at 100 chunks (Dense 0.840, NER3 0.800), NER3 passes Dense at 500 (0.670 vs.\ 0.660), and the margin then widens monotonically to 0.150 at full scale (0.600 vs.\ 0.450). This widening NER3-over-Dense gap is the one scale-dependent advantage the graph exhibits: against a dense bi-encoder---though not against BM25---topological structure degrades more gracefully than embedding similarity as the pool grows.

We read Figure~\ref{fig:scaling} as sharpening, not contradicting, the claim in Section~\ref{sec:open_corpus}: BM25 dominates from 500 chunks upward on this corpus, so the graph never overtakes the lexical baseline; the graph's genuine scale advantage is over dense retrieval, which it passes at 500 chunks and outperforms by a widening margin thereafter. The more striking scale effect, however, is the collapse of dense retrieval itself---strongest at small scale, weakest at production scale.

\subsection{Trade-Off Analysis}

BCT's closed-pool coverage gains come at a precision@1 cost ($-4.5$ points on 2WikiMultiHopQA, $-6.0$ on MuSiQue): over-seeding the personalization vector with aliases diffuses probability mass across more candidate bridge passages, trading top-1 precision for hit rate. This is desirable in closed pools, but as Table~\ref{tab:financial_eval} shows it becomes a liability at open-corpus scale, where the same diffusion spreads mass too thinly and precision falls below a lexical baseline.

\section{Conclusion and Future Work}
We presented Aethel and evaluated it across two regimes. In closed candidate pools, the coreference-aware BCT layer improves hit-rate coverage (HR@5 100.0\% on 2WikiMultiHopQA, 88.5\% on MuSiQue) at a precision@1 cost. At open-corpus scale over 4{,}123 real financial chunks, BM25 leads on HR@5; Reciprocal Rank Fusion with Aethel-NER3 ($k{=}60$, untuned) matches BM25's HR@1 and edges its MRR (0.479 vs.\ 0.468), but a paired bootstrap (95\% CI $[-0.163,+0.192]$, $N{=}20$) shows the gap does not clear zero. A scaling experiment (Figure~\ref{fig:scaling}) shows no graph/lexical crossover at any tested size and surfaces a sharper effect: dense retrieval collapses by 0.390 HR@5 as the corpus grows while BM25 degrades by only 0.100. The central finding is that graph-RAG's advantage is pool-size dependent and, at this scale, does not displace a strong lexical baseline; the most robust scale effect is the distractor-sensitivity of dense retrieval.

Two directions follow. First, \textbf{scale-aware seeding}: adaptive damping or seed-restriction to keep PPR mass from over-diffusing as the pool grows, potentially recovering the top-of-ranking signal fusion hints at. Second, \textbf{a domain-annotated PE diligence benchmark} at larger $N$ with multiple annotators, to move the open-corpus results from preliminary to conclusive.

\section*{Limitations}

While Aethel demonstrates a favorable coverage tradeoff in closed candidate pools, several limitations must be noted. First, the entity extraction and graph indexing phase relies on the disciplined taxonomy of institutional financial documents; less structured conversational inputs may introduce link noise into the entity-node mapping layer. Second, BCT's alias-expansion strategy improves hit-rate coverage but reduces precision@1, since over-seeding the PPR personalization vector can shift top-ranked results away from the single most relevant passage. Third, our closed-pool benchmarks use 200-question random samples from the official validation splits rather than the full test sets, which limits absolute comparability with full-scale system evaluations. Fourth, the BCT alias-expansion layer is currently implemented only in the standalone evaluator; integrating it into the production HippoRAG-backed pipeline is left as future work. Fifth, closed-pool retrieval metrics are computed over each question's provided context pool ($\sim$10--20 paragraphs) rather than an open corpus; these HR values reflect closed-set performance and, as Section~\ref{sec:open_corpus} demonstrates, are substantially lower in an open-retrieval setting. Sixth, the open-corpus financial evaluation uses 40 queries annotated by a single annotator; we therefore report effect sizes with bootstrap confidence intervals and treat these results as preliminary, and the fusion MRR difference in particular is not statistically distinguishable from zero at $N{=}20$. Seventh, an intermediate NER-graph variant contained a context-dependent adjacency defect that collapsed retrieval; the reported Aethel-NER3 results use a corrected substring-adjacency construction, and we disclose this to make the entity-index dependency explicit.

\clearpage
\appendix

\section{Prompt Templates and Swarm Agent System Directives}
\label{sec:appendix_prompts}
This section documents the formal execution prompts, schema constraints, and systemic context bounds assigned to each edge agent operating within the parallel multi-agent swarm architecture.

\subsection{Orchestrator Routing Node Micro-Logic}
The Orchestrator functions as a centralized task decomposition engine over the extracted sub-graphs. It coordinates agent state changes via the following target directive:
\begin{quote}\small
\texttt{SYSTEM CONTEXT: You are the Aethel System Routing Orchestrator operating inside an enterprise-grade financial compliance engine. Your task is to process highly fragmented, non-linear multi-hop financial queries over an input graph text collection. \\
INSTRUCTIONS: \\
1. Parse the localized entity-passage sub-graph returned by the power iteration algorithm. \\
2. Decompose the main query into separate, logically dependent analytical phases. \\
3. Assign each individual sub-task payload to a single domain expert node (Liquidity Specialist, Valuation Auditor, or Market Research Specialist). \\
4. Enforce structural boundaries: Agents must not cross-contaminate their memory contexts. You are solely responsible for gathering raw token extractions, initiating consensus validation loops, and formatting the verifiable cross-document audit trail. Respond strictly in structured JSON following the validation schema macro.}
\end{quote}

\subsection{Valuation Auditor Compliance Directive}
The Valuation Auditor isolates ledger accounts and recalculates asset statements to uncover internal or cross-document data conflicts:
\begin{quote}\small
\texttt{SYSTEM CONTEXT: You are the Aethel Valuation Auditor. You receive context vectors mapping out historical capital accounts, Net Asset Value (NAV) declarations, and distributed-to-paid-in (DPI) multiples isolated via Personalized PageRank iterations. \\
INSTRUCTIONS: \\
1. Extract numerical values associated with capital call balances and valuation ledger positions. \\
2. Re-compute mathematical aggregates across disjoint financial years to identify structural bookkeeping inconsistencies. \\
3. If a cross-document data discrepancy is detected where the variance exceeds a floating delta threshold of 0.01\%, flag the transaction as 'NON-COMPLIANT' and produce an explicit citation pointing to the conflicting text files. Do not guess or extrapolate.}
\end{quote}

\end{document}